\documentclass[twocolumn,amsmath,amssymb,pra,superscriptaddress]{revtex4-1}

\usepackage[final,letterspace=-40]{microtype}
\usepackage[colorlinks=true, linkcolor=blue, urlcolor=blue,  citecolor=blue, anchorcolor=blue]{hyperref}

\usepackage{mathpazo} 

\usepackage{graphicx}
\usepackage{dcolumn}
\usepackage{bm}
\usepackage{helvet}

\usepackage{caption}
\DeclareCaptionLabelFormat{plain}{Figure #2 $\boldsymbol |$ }

\begin{document}

\title{Coherence measurements of scattered incoherent light for \\lensless identification of an object's location and size}

\author{H. Esat Kondakci}
\affiliation{CREOL, The College of Optics \& Photonics, University of Central Florida, Orlando, Florida 32816, USA}
\author{Andre Beckus}
\affiliation{Department of Electrical and Computer Engineering, University of Central Florida, Orlando, FL 32816, U.S.A.}
\author{Ahmed El Halawany}
\author{Nafiseh Mohammadian}
\affiliation{CREOL, The College of Optics \& Photonics, University of Central Florida, Orlando, Florida 32816, USA}
\author{George K. Atia}
\affiliation{Department of Electrical and Computer Engineering, University of Central Florida, Orlando, FL 32816, U.S.A.}
\author{Ayman F. Abouraddy}\email{raddy@creol.ucf.edu}
\affiliation{CREOL, The College of Optics \& Photonics, University of Central Florida, Orlando, Florida 32816, USA}

\begin{abstract} \noindent 	In absence of a lens to form an image, incoherent or partially coherent light scattering off an obstructive or reflective object forms a broad intensity distribution in the far field with only feeble spatial features. We show here that measuring the complex spatial coherence function can help in the identification of the size and location of a one-dimensional object placed in the path of a partially coherent light source. The complex coherence function is measured in the far field through wavefront sampling, which is performed via dynamically reconfigurable slits implemented on a digital micromirror device (DMD). The impact of an object -- parameterized by size and location -- that either intercepts or reflects incoherent light is studied. The experimental results show that measuring the spatial coherence function as a function of the separation between two slits located symmetrically around the optical axis can identify the object transverse location and angle subtended from the detection plane (the ratio of the object width to the axial distance from the detector). The measurements are in good agreement with numerical simulations of a forward model based on Fresnel propagators. The rapid refresh rate of DMDs may enable real-time operation of such a lensless coherency imaging scheme.   
\end{abstract}

\small
\maketitle

\section{Introduction}
\noindent 
When incoherent or partially coherent light scatters off an obstructive object, the shadow formed in the vicinity of the object gradually blurs at larger distances until the scattered field ultimately exhibits a smooth distribution with only feeble local intensity variations. In absence of a lens to form an image, it is difficult to reconstruct the scattering object from a measurement of the far-field intensity alone. Although image processing can help improve the quality of a recorded image by removing blur resulting from motion or poor focusing \cite{Huang79Book,Jain89Book,Jahne97Book}, it remains a notoriously difficult task to undo the blurring from diffractive spreading after free propagation. Although the transfer function for free propagation of incoherent light does \textit{not} include zeros (for an infinitely sized detector), the decay of the transfer function with spatial frequency is nevertheless extremely sharp \cite{George97OC}, which makes the inversion sensitive to noise. In other words, the remnant spatial variations in the lensless far-field intensity distribution are too small to allow for object reconstruction. Other approaches to reconstruct a scattering object make use of phase retrieval \cite{Fienup81AO} with the measured intensity distributions in two planes \cite{Gerchberg72Optik}, or the amplitude \cite{Fienup78OL} or phase \cite{Hayes80IEEE,Hayes82OE} of the Fourier transform of the field -- with the phase information typically yielding better reconstructions \cite{Kermisch70JOSA,Huang75IEEE,Oppenheim81IEEE}. These approaches are usually more successful in object reconstruction when coherent light is used \cite{Zhang:03,Pedrini:05,Abouraddy06NM,Sorin09NL}.

We consider here an alternative for \textit{far-field lensless identification} of an object illuminated with incoherent or partially coherent light: instead of measuring the far-field \textit{intensity} profile, we measure the \textit{spatial coherence function} representing the correlation between pairs of points in the field. The problem is characterized fundamentally by two spatial scales: the transverse extent of the scattered-field \textit{intensity distribution} and the width of the \textit{spatial coherence function} associated with the field. In a lensless configuration, diffractive spreading can render the extent of the former quite large and devoid of distinctive features. The width of the spatial coherence function, however, may be considerably smaller and retain sufficient information to identify a scattering object. We consider thin planar one-dimensional (1D) objects whether obstructive or reflective -- with the other transverse dimension assumed uniform. Such an object is parametrized by three quantities: size, transverse position with respect to the optical axis, and longitudinal position with respect to the detection plane. In such a scenario, we find that measuring the complex spatial coherence (both magnitude and phase) enables object identification. Furthermore, even though the field is spread spatially over a large area, we need to sample only a limited spatial extent of the scattered field -- on the order of the transverse coherence length of the field at the detection plane. Surprisingly, the spatial extent of the required measurement in some situations may be \textit{smaller} than the physical size of the object itself, which could be located a large distance away from the detection plane.

There are many approaches for assessing the spatial coherence of an optical field. It was recognized long ago by Zernike that the visibility of Young's double slit interference \cite{Young04PTRS} reveals the spatial cross-coherence between the fields at the two locations of the slits \cite{Zernike38Physica}. Despite initial realizations of such measurements \cite{ThompsonPartiallyCoherent} (see Ref. \cite{Francon67PinO} for a review of early efforts), the tediousness involved in mapping out the coherence function by varying the double-slit separation has led to the development of a host of alternative approaches. For example, a lateral-shearing Sagnac interferometer can measure the spatial coherence function \cite{Iaconis96OL,Cheng00JMO}. Alternatively, a fixed double-slit separation can be exploited by laterally shifting the slits across the input field along with a reversed copy of the field created by a cube beam splitter that is also shifted laterally, a so-called `reversed-wavefront' Young interferometer \cite{Santarsiero06OL}. Other approaches make use of non-redundant arrays of apertures to multiplex interferograms \cite{Mejia07OC,Gonzales11JOSAA}, a pair of non-parallel slits for multiplexing one-dimensional interferograms \cite{DivittNonParallelSlits,DivittSunlight2015}, or exploit wavefront sensors \cite{Stoklasa14NC}. Another strategy for acquiring the spatial coherence function relies on phase-space methods that exploit the connection between spatial coherence and the Wigner distribution associated with the field \cite{Cho12OL,Wood14OL,Sharma16OE}.

\begin{figure*}[t!]
\centering
\includegraphics[scale=1]{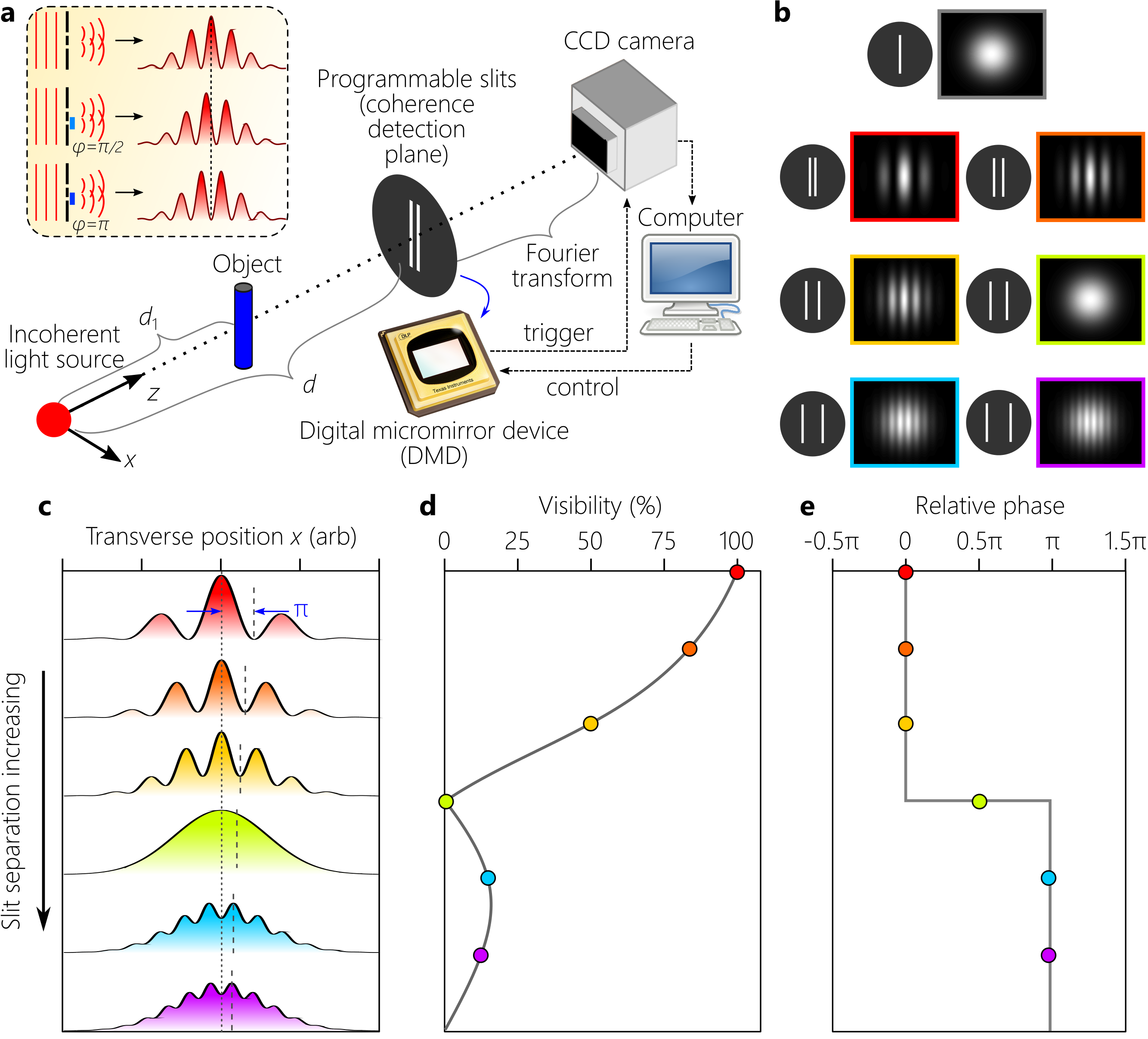}
  \caption{Complex-coherence measurements of incoherent light scattered from an object. (a) Schematic of the experimental setup. Incoherent light (an LED) is incident on a 1D object and the scattered light impinges on computer-controlled double slits realized via a digital micromirror device (DMD). We depict the object obstructing the field (we also perform the experiment with only the reflected light reaching the detector). A spatial Fourier transform is implemented between the double slits and the CCD camera. In the experiment, the slits are realized in reflection mode, and is followed by a pair of lenses for optical relay and a cylindrical lens that produces the Fourier transform (omitted from the schematic for simplicity). The inset depicts the effect of the phase $\varphi\!=\!\mathrm{Arg}\{G(x_{1}\!-\!x_{2})\}$ on the interference pattern. When $\varphi\!=\!\pi$, the intensity at the center of the interferogram is a minimum. (b) Interference patterns produced at the CCD while increasing the slit separation. The period of the interferogram decreases with increased separation, and the visibility drops with potential revivals at larger separations. When only one slit is activated, a diffraction pattern with no interference is registered (top panel). (c) 1D sections $I(x)$ through interferograms (vertically displaced for clarity) with decreasing period as the slit separation is increased. The phase $\mathrm{Arg}\{G(x_{1}\!-\!x_{2})\}$ is measured relative to the dotted vertical reference. The shifts between the dotted and dashed lines corresponds to a $\pi$-phase. (d) From each interferogram in (c) we obtain the visibility $V$ as a function of slit separation (each colored circle is extracted from one interferogram). The visibility is related to the amplitude of the degree of coherence $|g(x_{1}\!-\!x_{2})|\!=\!V(I_{1}\!+\!I_{2})/(2\sqrt{I_{1}I_{2}})$. (e) The phases $\mathrm{Arg}\{G(x_{1}\!-\!x_{2})\}$ are extracted from the displacement of the local maxima and minima around the center fringe, with phase 0 assigned to the first interferogram.}
\label{Fig1:Concept}\vspace*{-0.5cm}
\end{figure*}

The time-consuming aspect of direct wavefront sampling to map the spatial coherence function has been recently obviated with the emergence of digital micromirror devices (DMD), also known as digital light processors (DLP) \cite{Dudley03SPIE}. These devices are formed of a two-dimensional arrays of micromirrors that can be individually addressed to direct light into one of two directions, and thus can be used as a pixellated on/off modulator for the optical wavefront. Such devices operate at considerably higher rates than liquid-crystal-based spatial light modulators \cite{Tripathi14OE}, and have been used in introducing photonic time delays \cite{Riza97SPIE}, beam profiling \cite{Sheikh09PTL}, synthesizing special optical beams \cite{Ren15AP}, and wavefront-splitting interferometry \cite{Torre16SPIE}. Most recently, these devices have been utilized in mapping out the coherence function by implementing dynamically reconfigurable pairs of amplitude slits for measurements using coherent \cite{Vdovin15JO} and incoherent \cite{CoherenceDMD} light.

We measure the complex coherence function here via dynamically configured double slits implemented with a DMD. The amplitude and phase of the coherence function at any pair of points $(x_{1},x_{2})$ are assessed from the fringe \textit{visibility} and fringe \textit{shift}, respectively, observed in the interferogram produced by two slits placed at positions $x_{1}$ and $x_{2}$. The high refresh rate of the DMD allows slit patterns to be quickly cycled. Our work aims at recovering characteristic features of an object in a lensless configuration through an examination of the coherence function produced by scattered partially coherent quasi-monochromatic light over a small area in the far field. The source in our experiments is an extended-area LED whose partial coherence is represented by a truncated Gauss-Schell model \cite{Schell67IEEE}. Whereas measurements of the coherence function of various light sources are well-documented \cite{ThompsonPartiallyCoherent,Marks99AO,CoherenceDMD,DivittSunlight2015}, here we focus on the deviations in the measured coherence with respect to that of the source as introduced by an object lying in the field's path. We consider both \textit{intercepting} (obstructing) objects placed in front of a light source to block part of the beam, as well as \textit{reflecting} objects placed such that they reflect light toward the detector. In all cases, we compare the measurements to theoretical predictions obtained using Fresnel propagators.

The paper is organized as follows. We present the theoretical framework in Section~\ref{Section:Method}, where we also describe the measurement setup and the algorithm used for extracting the coherence magnitude and phase. In Section~\ref{Section:Results}, we discuss the experimental results for obstructive and reflective objects and compare them with theoretical predictions. In the conclusions we consider prospects for extending this methodology to more complex `scenes' comprising multiple objects at different locations.

\section{Coherence Function for Partially Coherent Light Scattering off an Object} \label{Section:Method}

\subsection{Theoretical method}

Our overall strategy is depicted in Fig.~\ref{Fig1:Concept}(a). In these experiments, we measure the degree of spatial coherence
$g(x_{1},x_{2};\lambda)$ (also known as the spectral degree of coherence) at two points with coordinates $x_{1}$ and $x_{2}$ in a quasi-monochromatic scalar field, where $\lambda$ is the wavelength. Here $g(x_{1},x_{2};\lambda)$ is related to $G(x_{1},x_{2};\lambda)\!=\!\langle E(x_{1})E^{*}(x_{2})\rangle$ through a normalization with respect to the intensity, $g(x_{1},x_{2})\!=\!G(x_{1},x_{2})/\sqrt{I(x_{1})I(x_{2})}$; where $E(x)$ is a realization of the electric field, $I(x_{1})\!=\!G(x_{1},x_{1})$, $\langle\cdot\rangle$ denotes statistical averaging over an ensemble, and we drop the wavelength dependence henceforth for simplicity. We restrict ourselves to field variations along a single transverse coordinate $x$ (assuming all fields are uniform along the $y$ coordinate).

The field propagating from the source plane to the detection plane (a total distance of $d$) undergoes a mapping through a linear system represented by an impulse response function $h(x_{1},x')$, where $x_{1}$ is a point in the detection plane and $x'$ is a point in the source plane. In our experimental arrangement, this system consists of a succession of three linear sub-systems: (1) free-space propagation a distance $d_{1}$ from the source plane $x'$ to the object plane $\tilde{x}$; (2) transmission or reflection from an object located at the plane $\tilde{x}$; and (3) free-space propagation a distance $d_{2}$ from the object plane $\tilde{x}$ to the detection plane $x_{1}$. Free propagation a distance $d$ in the Fresnel regime is described by the impulse response function 
\begin{equation}
h_{\mathrm{F}}(x_{1},x';d)=\frac{\exp(ikd)}{\sqrt{i\lambda d}}\exp\left\{i\frac{k}{2d}(x_{1}-x')^2\right\},
\end{equation}
where $k\!=\!{2\pi}/\lambda$ is the wavenumber \cite{FundOfPhotonics}.

The object is assumed to be thin and described by a real-valued transmittance function $t(x)$ -- although this model readily accommodates a complex-valued transmittance. Both intercepting (obstructing) and reflecting objects are modeled -- for simplicity -- as indicator functions; that is, light at any point in the object plane either passes unobstructed, or is blocked completely. For an intercepting object, $t(x)\!=\!1\!-\!\mathrm{rect}\left(\frac{x-x_{0}}{w}\right)$, where $x_{0}$ and $w$ are the object center position with respect to the optical axis and its width, respectively, and $\mathrm{rect}(x)\!=\!1$ when $-0.5\!\leq\!x\!\leq\!0.5$ and is zero otherwise. The reflective object is assumed to be specular, and so is modeled as an aperture with transmission function $t(x)\!=\!\mathrm{rect}\left(\frac{x-x_{0}}{w}\right)$. Objects with transmittance of values other than 0 or 1 can also be accommodated within this framework. The object is thus identified by three parameters: its width $w$; its transverse position $x_{0}$; and its axial distance from the detection plane $d_{2}$ (for fixed total distance from source to detector $d$).

The impulse response function $h(x_{1},x')$ of the entire system from the source to detector is therefore given by
\begin{equation}\label{Eq:TotalSystem}
h(x_{1},x')=\int\!\!d\tilde{x}\,h_{\mathrm{F}}(x_{1},\tilde{x};d_{2})\;t(\tilde{x})\;h_{\mathrm{F}}(\tilde{x},x';d_{1}).
\end{equation}
The coherence function at a pair of points $x'$ and $x''$ in the \textit{source} plane of $G_{\mathrm{s}}(x',x'')$ is mapped to a pair of points $x_{1}$ and $x_{2}$ in the \textit{detection} plane of $G(x_{1},x_{2})$ via the transformation
\begin{equation}\label{Eq:InputToOutputCoherence}
G(x_{1},x_{2})=\!\int\!\!\!\!\int\!dx'dx''h(x_{1},x')\;h^{*}(x_{2},x'')\;G_{\mathrm{s}}(x',x''). 
\end{equation}
Using this forward model, the coherence at the detector plane can be evaluated once the source is known, which requires a reference measurement. Finally, the calculation results are integrated over the source spectral bandwidth (1~nm in our experiments) \cite{OpticalCoherenceBook}.

Although we are considering 1D fields (one transverse coordinate $x_{1}$), the associated coherence function $G(x_{1},x_{2})$ is a 2D distribution. From $G(x_{1},x_{2})$ we  measure only its anti-diagonal, i.e., as a function of the \textit{separation} $x_{1}\!-\!x_{2}$ between pairs of points located symmetrically around the optical axis $x_{1}\!=\!0$; the intensity is the diagonal $I(x_{1})\!=\!G(x_{1},x_{2})$. Hereafter, we adopt the notation $G(x_{1}\!-\!x_{2})$ to emphasize that the measured data is a 1D function.

\subsection{Measurement system}

We measure $g(x_{1}\!-\!x_{2})$ using the system illustrated in Fig.~\ref{Fig1:Concept}(a), which involves implementing parallel double slits via a DMD (Texas Instruments DLP6500, $1920\!\times\!1080$ pixels with a pitch of $\approx\!7.56$~$\mu$m). The width of each slit is 3~pixels, $\approx\!22.7$~$\mu$m. The source is a spatially extended LED (Thorlabs, M625L3) with a peak wavelength of $\approx\!633$~nm and a FWHM-bandwidth of $\approx\!18$~nm that is spectrally filtered by a $\approx\!1.3$-nm-FWHM band-pass filter centered at 632.8~nm (Thorlabs, FL632.8-1). Partially coherent light reflected from the double slits on the DMD creates an interference pattern whose fringe period decreases with increasing slit separation [Fig.~\ref{Fig1:Concept}(b,c)], which we vary in the range 0 to 1~mm while remaining centered at the optical axis. Light reflected from the DMD is first relayed by an imaging system formed of two lenses of focal lengths 10 and 20~cm providing $2\times$ magnification, followed by a cylindrical lens of focal length $f\!=\!20$~cm in a $2f$ configuration that produces the interference fringes, which are recorded by a CCD camera (The ImagingSource, DFK 31BU03). The system is fully automated: the refresh rate for displaying the double slits on the DMD with different separations $x_{1}\!-\!x_{2}$ is synchronized with the exposure and acquisition time of the CCD.

\subsection{Extracting the coherence function magnitude and phase}

The interference patterns recorded by the CCD take the form 
\begin{widetext}
\begin{equation}\label{Eq:Interferogram}
I(x)\propto\mathrm{sinc}^{2}\left(\frac{kx\ell}{2\pi Mf}\right)\left\{I_{1}+I_{2}+2|G(x_{1}\!-\!x_{2})|\cos{\left(\frac{kx}{Mf}(x_{1}-x_{2})-\varphi\right)}\right\},
\end{equation}
\end{widetext}
where $\varphi\!=\!\mathrm{Arg}\{G(x_{1}\!-\!x_{2})\}$, $\ell\!\approx\!22.7$~$\mu$m is the slit width, and $I_{1}$ and $I_{2}$ are the peak values of the diffraction patterns from each slit, which can be obtained by activating one slit at a time, $\mathrm{sinc}(x)\!=\!\tfrac{\sin{(\pi x)}}{\pi x}$, and $M\!=\!2$ is the magnification of the optical relay preceding the $2f$ Fourier transform system comprising a lens of focal length $f\!=\!20$~cm. We obtain $|g(x_{1}\!-\!x_{2})|$ from the visibility $V$ of the recorded interferograms [Fig.~\ref{Fig1:Concept}(d)] along with intensity measurements from individual slits, whereas the phase $\mathrm{Arg}\{g(x_{1}\!-\!x_{2})\}$ from the displacement of the central fringe with respect to a reference [Fig.~\ref{Fig1:Concept}(e)].

\begin{figure*}[t!]
\centering
\includegraphics[scale=1]{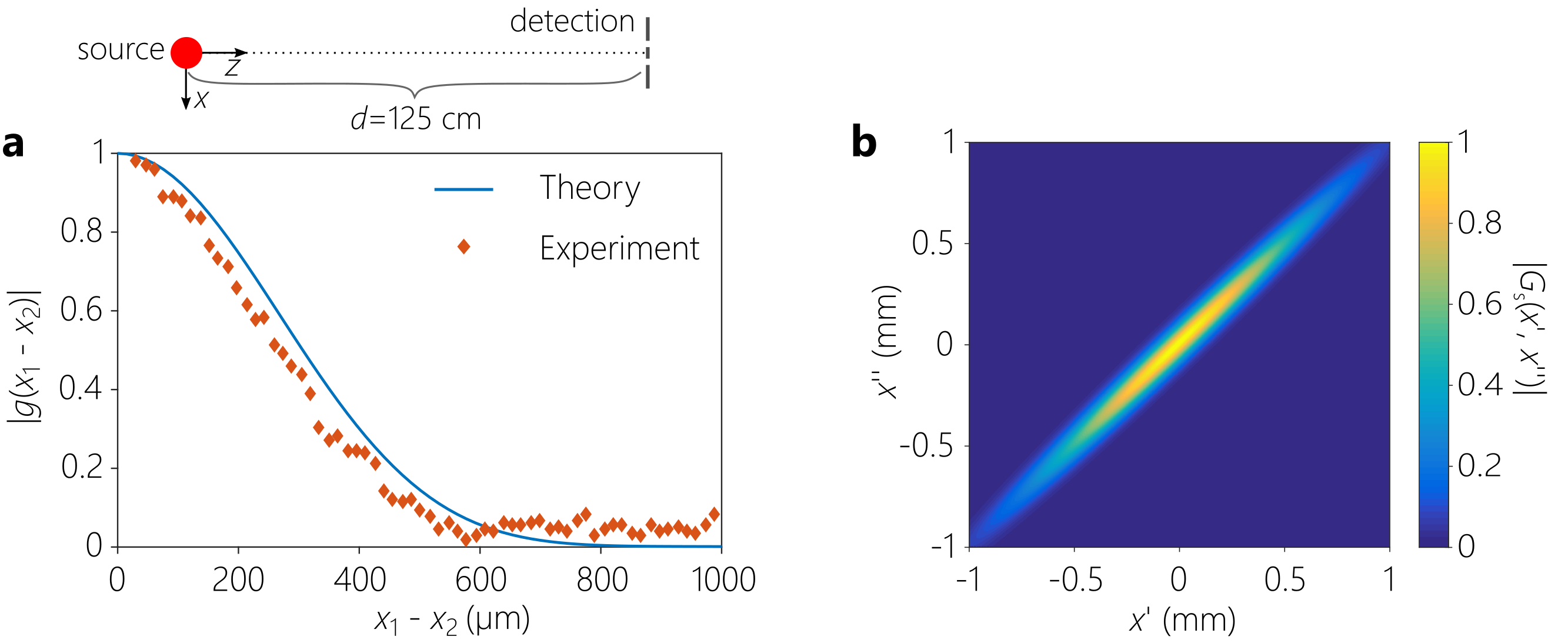}
\caption{(a) The measured magnitude of the degree of spatial coherence $|g(x_{1}-x_{2})|$ (diamonds) at the detector plane in absence of an object (uninterrupted propagation from the source to the detector). The solid line is a theoretical fit. The measured and theoretical phase $\mathrm{Arg}\{g(x_{1},x_{2})\}$ is zero over this range [see Fig.~\ref{Fig3:varying-width}(b)]. The schematic above the panel depicts the measurement geometry. The red circle is the LED source and the DMD implementing the double slits is placed at the detection plane. (b) A plot of the spatial coherence function magnitude of the source $G_{\mathrm{s}}(x',x'')$ based on Eq.~\ref{Eq:SourceGaussSchellModel} that makes use of the parameters extracted from the measurements in (a).}
\label{Fig2:source-char}
\end{figure*}

The visibility $V\!=\!(I_{\mathrm{max}}-I_{\mathrm{min}})/(I_{\mathrm{max}}+I_{\mathrm{min}})$ is obtained from the measured interferogram, 
where $I_{\mathrm{max}}$ and $I_{\mathrm{min}}$ are the maximum and minimum values of $I(x)$, respectively, from which we obtain $|g(x_{1}\!-\!x_{2})|\!=\!\frac{I_{1}+I_{2}}{2\sqrt{I_{1}I_{2}}}V$ [Fig.~\ref{Fig1:Concept}(b-d)]. To extract the phase $\mathrm{Arg}\{g(x_{1}\!-\!x_{2})\}$, we estimate the displacement of the interference patterns at different separations $x_{1}-x_{2}$ with respect to a fixed fringe location. The shift in the location of the central fringe with respect to the reference gives the phase $\mathrm{Arg}\{g(x_{1}\!-\!x_{2}\}$ once the shift is normalized with respect to the fringe period [Fig.~\ref{Fig1:Concept}(e)]. 

Care must be taken with respect to the \textit{maximum} coherence width (slit separation) at the detection plane that can be measured by this system. When individual pixels of the DMD are activated, they tilt approximately 12.5$^{\circ}$ away from the normal to the DMD plane, thus creating an angle $\psi\!\approx\!25^o$ between the incident and reflected fields. This feature of the DMD pixel architecture introduces a longitudinal path length difference of $\Delta z\!=\!(x_{1}\!-\!x_{2})\sin{2\psi}$ between light reflected from positions $x_{1}$ and $x_{2}$. This extra path length of course increases with increased slit separation. The finite bandwidth of the radiation therefore results in a gradual loss of coherence between the fields at $x_{1}$ and $x_{2}$ with increased separation $x_{1}\!-\!x_{2}$ even if the fields are \textit{spatially coherent}. We compensate for this artifact of the measurement scheme by introducing a premultiplier to the measured $g(x_{1}\!-\!x_{2})$. Assuming a Gaussian spectral profile, this premuliplier takes the form of an inverted Gaussian of 627-$\mu$m-FWHM. We found that reducing this value by 15\% to 533~$\mu$m offers an excellent match between the theoretical predictions based on the model presented in the previous subsection and all the measurements. We attribute this discrepancy to the deviation of the actual spectral linewidth of the radiation from the presumed Gaussian form.

\section{Experimental Results} \label{Section:Results}

\subsection{Source Characterization}

We first carry out a reference measurement -- in absence of any object -- to characterize the light source. To capture the characteristics of the partial coherence of the source, we make use of a Gauss-Schell model \cite{Schell67IEEE,Deschamps83JOSA,He98OC} in which a jointly Gaussian coherence function (along the $x'\!+\!x''$ and $x'\!-\!x''$ directions) has its intensity profile truncated. The intensity of the source is taken to be Gaussian but is spatially limited by a width equal to the size of the LED ($\approx2$~mm). The Gauss-Schell model is parameterized by the beam width $\alpha$, the spatial coherence width $\sigma$, and aperture width $L$. The source is quasi-monochromatic modeled with a uniform spectral profile having a center wavelength $\lambda_{0}\!=\!633$~nm and bandwidth $\Delta\lambda\!=\!1.3$~nm. Therefore, the full coherence function of the source is given by
\begin{widetext}
\begin{equation}\label{Eq:SourceGaussSchellModel}
G_{\mathrm{s}}(x',x'';\lambda)=\exp\left(-\frac{(x'+x'')^2}{2 \alpha^2}\right)\exp\left(-\frac{(x'-x'')^2}{2\sigma^2}\right)
\mathrm{rect}\left(\frac{x'}{L}\right)\mathrm{rect}\left(\frac{x''}{L}\right)\mathrm{rect}\left(\frac{\lambda-\lambda_{0}}{\Delta\lambda}\right).
\end{equation}
\end{widetext}

We plot in Fig.~\ref{Fig2:source-char}(a) the measured magnitude $|g(x_{1}\!-\!x_{2})|$. We fit the measurements to theoretical predictions based on propagating the source Gauss-Schell model in Eq.~\ref{Eq:SourceGaussSchellModel} to the detector plane unimpeded (no obstructing object) via Eq.~\ref{Eq:InputToOutputCoherence} after setting $h(x_{1},x)\!=\!h_{\mathrm{F}}(x_{1},x;d)$ with $d\!=\!125$~cm. From the fitting procedure, we estimate the remaining parameters $\alpha$ and $\sigma$ of the source to be $\alpha\!=\!1\!/\!\sqrt{2\ln2}$~mm and $\sigma\!=\!75\!/\!\sqrt{2\ln2}$~$\mu$m, which yield a FWHM beam width of 1~mm and a FWHM coherence width of 75~$\mu$m. The model for the source coherence function utilizing these parameters is given in Fig.~\ref{Fig2:source-char}(b). This reference measurement that enabled us to reconstruct the source coherence function $G_{\mathrm{s}}(x',x'')$ is used subsequently in Eq.~\ref{Eq:InputToOutputCoherence} once an object is placed in the field's path.

\begin{figure*}[t!]
\centering
\includegraphics[scale=1]{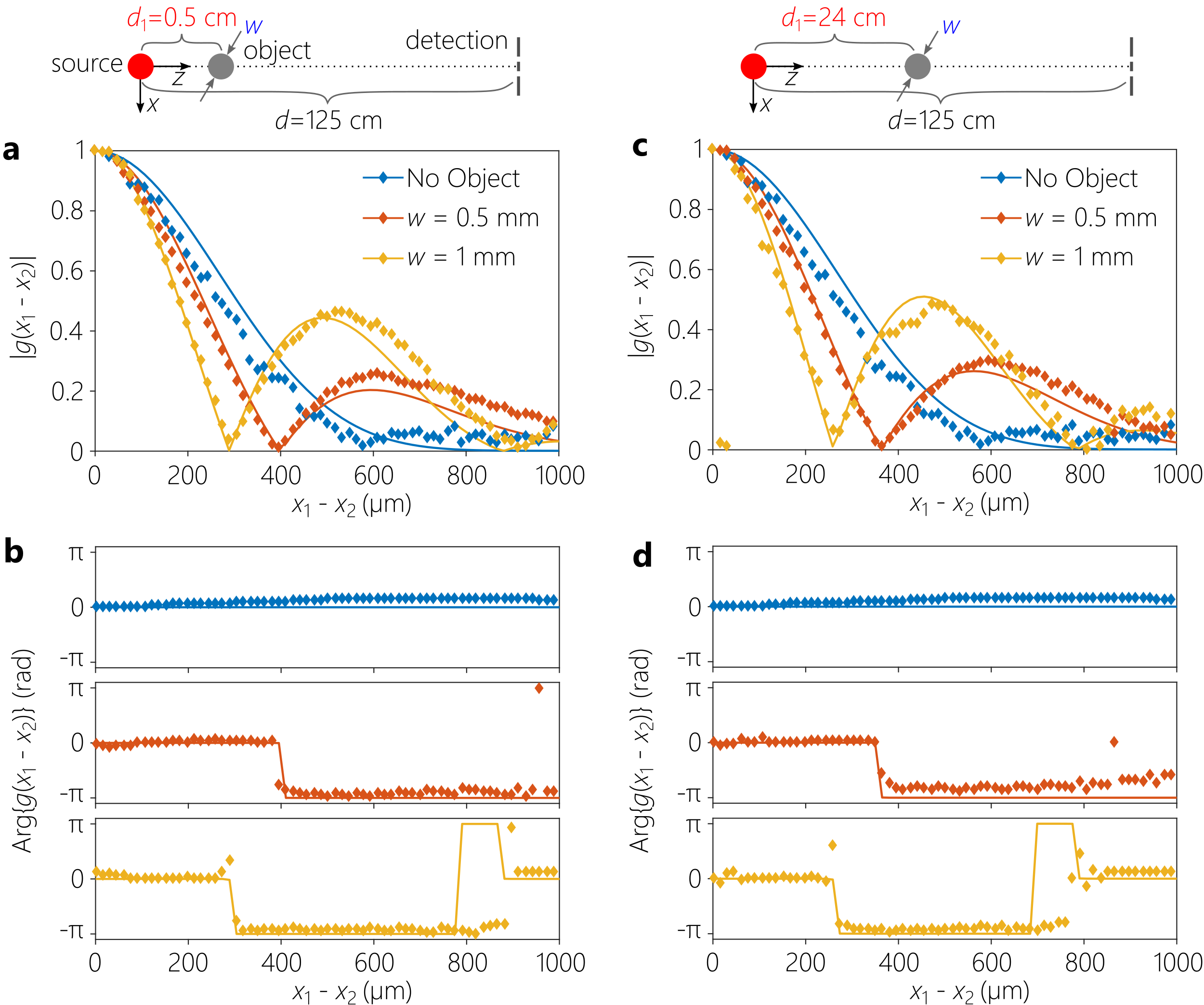}
\caption{Impact of the object size $w$ on the degree of spatial coherence $g(x_{1}\!-\!x_{2})$ when the object location $(x_{0},d_{1})$ is held fixed. (a) The measured magnitude of the degree of coherence $|g(x_{1}\!-\!x_{2})|$ in three cases: unimpeded propagation from the source to the detection plane (no object, $w\!\rightarrow\!0$), $w\!=\!0.5$~mm, and $w\!=\!1$~mm. In all cases $x_{0}\!=\!0$, $d_{1}\!=\!0.5$~cm, and $d\!=\!125$~cm. The schematic above the panel depicts the measurement geometry (the gray circle is the object). The data points are plotted as diamonds, and the solid lines are theoretical predictions based on Eqs.~\ref{Eq:TotalSystem},~\ref{Eq:InputToOutputCoherence}, and~\ref{Eq:SourceGaussSchellModel}. (b) The measured phases $\mathrm{Arg}\{g(x_{1}\!-\!x_{2})\}$ corresponding to the three cases plotted in (a). The diamonds are data points and the solid lines are theoretical predictions. (c-d) Same as (a-b) except that $d_{1}\!=\!24$~cm; that is, the object is placed farther away from the source and closer to the detection plane (the total distance from source to the DMD is held fixed at $d\!=\!125$~cm).}
\label{Fig3:varying-width}
\end{figure*}

\subsection{Coherence function due to intercepting objects}

We first consider an intercepting object in the form of a thin metal wire of diameter $w\!=\!0.5$~mm or 1~mm placed between the light source and the detector such that it partially blocks light from reaching the detector. We consider locating the object at different axial distances from the source ($d_{1}\!=\!0.5$, 24, and 72~cm) and at various positions along the transverse plane ($x_{0}\!=\!0$, $\pm50$, and $\pm100$~$\mu$m with respect to the optical axis). In each experiment we hold two of these parameters fixed while varying the third. Because of the small size of the object ($\leq\!1$~mm) placed in an incoherent field and the large distance to the detection plane ($\sim\!1$~m), the intensity distribution at the detection plane (DMD) does not display a clear shadow or directly indicate the existence of an object. Instead, a flat intensity profile is observed over the DMD ($\sim\!1$-mm width under consideration). We proceed to show that measuring the two-point field correlations -- encoded in $g(x_{1}\!-\!x_{2})$ -- over this same spatial extent can help identify the object. Note that $g(x_{1}\!-\!x_{2})\!=\!g^{*}(x_{2}\!-\!x_{1})$, so we plot $g(x_{1}\!-\!x_{2})$ for $x_{1}\!-\!x_{2}\!\geq\!0$ only.

\subsubsection{Impact of the object size}

We first examine the effect of the object size $w$ -- when its location $(x_{0},d_{1})$ is fixed -- on the coherence function $g(x_{1}\!-\!x_{2})$ in Fig.~\ref{Fig3:varying-width}. The presence of the object reduces the width of the main lobe of $|g(x_{1}\!-\!x_{2})|$ and introduces a significant side lobe, features that did not exist in the source coherence function measured at the detection plane in absence of an object [Fig.~\ref{Fig2:source-char}(a)]. Increasing the object width increases the side-lobe peak amplitude and reduces the width of the lobes [Fig.~\ref{Fig3:varying-width}(a)].

This can be understood by realizing that the obstructing object modulates the field intensity at the object plane $\tilde{x}$, which now represents a secondary source. In the far-field, the van Cittert-Zernike theorem indicates that the distribution of spatial coherence is related to the Fourier transform of this secondary source intensity distribution when the field is incoherent \cite{FundOfPhotonics}. The general trends dictated by the van Cittert-Zernike therorem still apply when the field is partially coherent, as is our case here. The nulls of $|g(x_{1}\!-\!x_{2})|$ remain associated with abrupt jumps in phase by $\pi$ [Fig.~\ref{Fig3:varying-width}(b)]. Similar results are observed when the object approaches the detection plane [Fig.~\ref{Fig3:varying-width}(c-d)], with the nulls in $|g(x_{1}\!-\!x_{2})|$ occurring at smaller values of $x_{1}\!-\!x_{2}$.

\begin{figure*}[t!]
\centering
\includegraphics[scale=1]{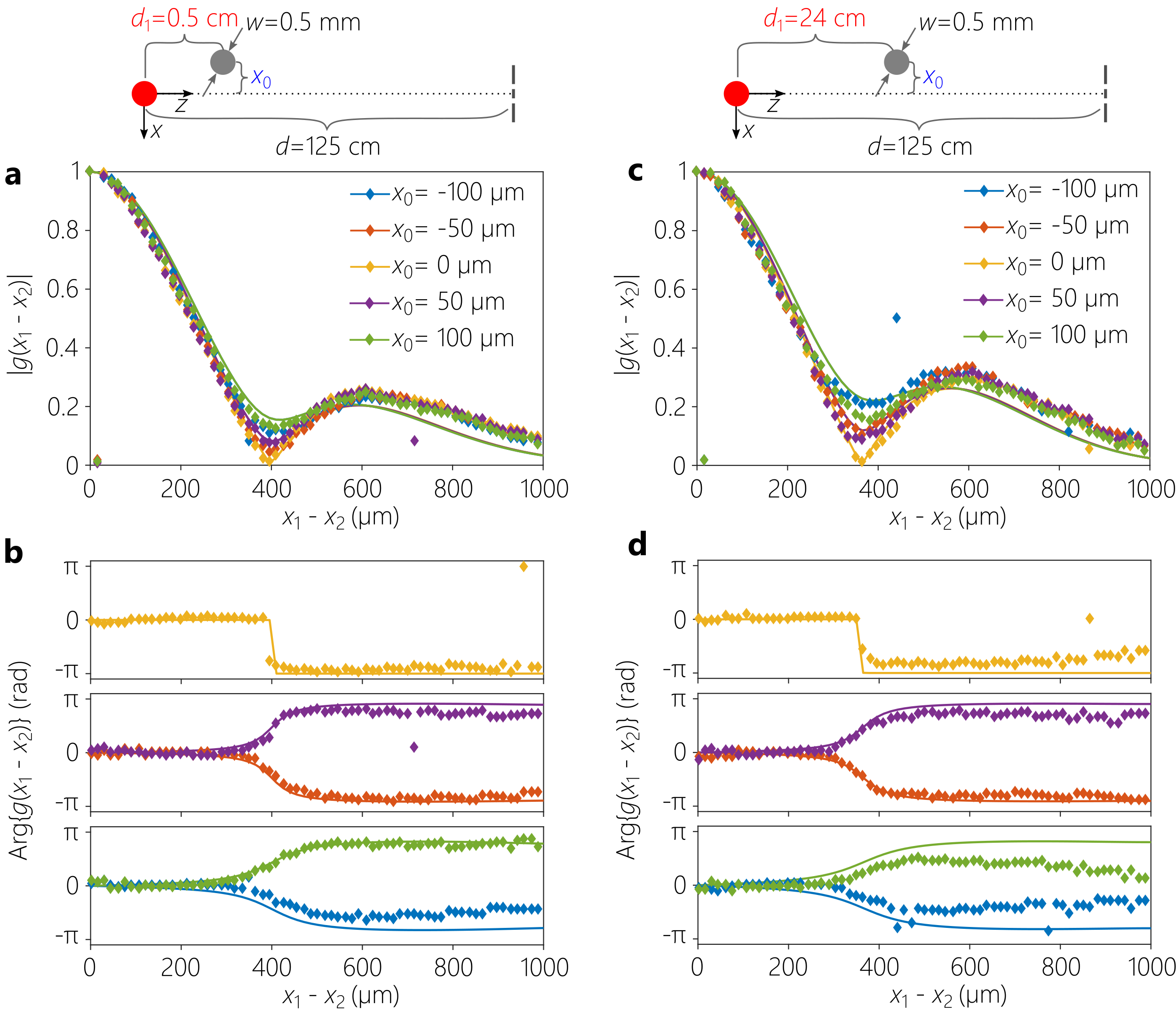}
\caption{Comparison of degree of spatial coherence for as the object is displaced in the transverse plane with respect to the optical axis. (a) Coherence magnitude $|g(x_{1}\!-\!x_{2})|$ and (b) phase $\mathrm{Arg}\{g(x_{1}\!-\!x_{2})\}$ are shown for an object placed $d_{1}\!=\!0.5$~cm while varying $x_{0}$ from $-100$~$\mu$m to $100$~$\mu$m. (c-d) Same as (a-b) except that the object is placed at $d_{1}\!=\!24$~cm from the source. The object width is $w\!=\!0.5$~mm and the total distance from source to the detection plane is $d\!=\!125$~cm.}
\label{Fig4:varying-x0}
\end{figure*}

\subsubsection{Impact of the object transverse location}

Figure~\ref{Fig4:varying-x0} shows the impact of changing the transverse position $x_{0}$ of an object of fixed size ($w\!=\!0.5$~mm) moving in a plane at a fixed distance from the detector. When the object is located on the optical axis, a zero is observed in the coherence function $g(x_{1}\!-\!x_{2})\!=\!0$ for some value of $x_{1}\!-\!x_{2}$ set by the object size [Fig.~\ref{Fig4:varying-x0}(a)]. At this null, the phase $\mathrm{Arg}\{g(x_{1}\!-\!x_{2})\}$ undergoes an abrupt jump of $\pi$. As the object moves away from the optical axis, the coherence function does not reach zero at the first minimum. Additionally, in lieu of the abrupt $\pi$-phase jump, a gradual transition in phase takes place [Fig.~\ref{Fig4:varying-x0}(b)]. As the object moves further away from the optical axis, the drop in $|g(x_{1}\!-\!x_{2})|$ at the first minimum is further diminished and the associated phase change becomes even more gradual.

A measurement of $|g(x_{1}\!-\!x_{2})|$ alone results in an inherent ambiguity with respect to the \textit{direction} of displacement of the object with respect to the optical axis. Therefore the measurements and theoretical predictions for $|g(x_{1}\!-\!x_{2})|$ coincide for displacements of $\pm x_{0}$. This ambiguity is lifted by observing the \textit{phase} $\mathrm{Arg}\{g(x_{1}\!-\!x_{2})\}$. The gradual phase change at the first minimum of $|g(x_{1}\!-\!x_{2})|$ is in opposite directions for the positive and negative values of $x_{0}$, thus helping to identify the object location. Furthermore, both effects that result from a transverse displacement -- lifting of the zeros of $g(x_{1}\!-\!x_{2})$ and gradual change in $\mathrm{Arg}\{g(x_{1}\!-\!x_{2})\}$ -- are further enhanced as the object approaches the detection plane [Fig.~\ref{Fig4:varying-x0}(c,d)].

\begin{figure*}[t!]
\centering
\includegraphics[scale=1]{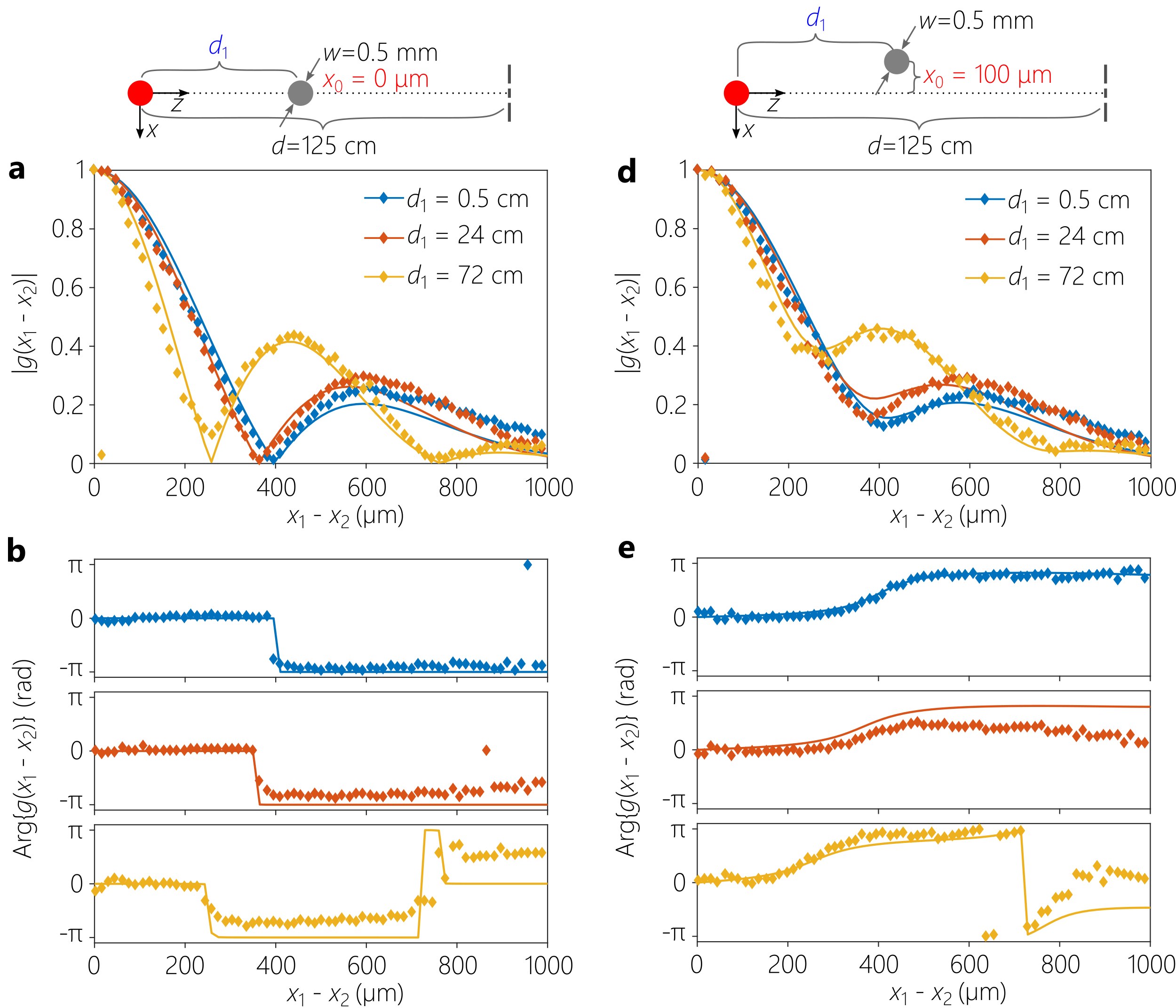}
\caption{Comparison of degree of spatial coherence for various source-to-object axial distances $d_{1}$. (a) Coherence magnitude $|g(x_{1}\!-\!x_{2})|$ and (b) phase $\mathrm{Arg}\{g(x_{1}\!-\!x_{2})\}$ are shown for an object placed on the optical axis $x_{0}\!=\!0$. (c-d) Same as (a-b) except that the object is displaced from the optical axis to $x_{0}\!=\!100$~$\mu$m. The object width is $w\!=\!0.5$~mm and the total distance from source to the detection plane is $d\!=\!1.25$~m.}
\label{Fig5:varying-d1}
\end{figure*}

\begin{figure*}[t!]
\centering
\includegraphics[scale=1]{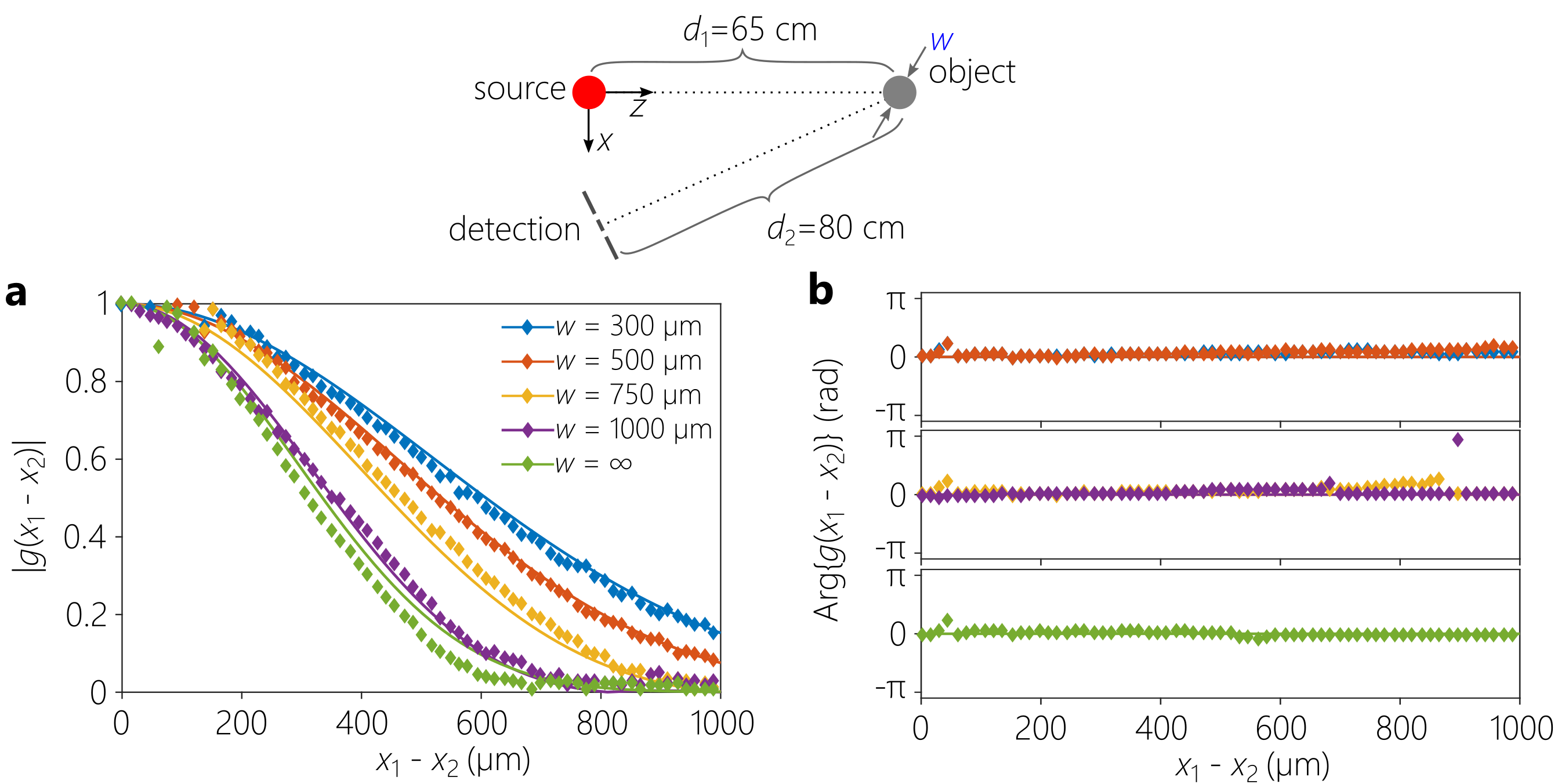}
\caption{Comparison of the degree of spatial coherence for reflective objects of varying widths $w$. The distance from the source to the object is $d_{1}\!=\!65$~cm, and the total distance from source to DMD is $d\!=\!1.45$~m. The objects are placed on the optical axis defines by the source $x_{0}\!=\!0$. (a) The coherence magnitude $|g(x_{1}\!-\!x_{2})|$ and (b) phase $\mathrm{Arg}\{g(x_{1}\!-\!x_{2})\}$ are shown while varying $w$. Experimental results are plotted with diamonds and theoretical predictions are solid lines. The infinite width case is equivalent to free space propagation. The phase $\mathrm{Arg}\{g(x_{1}\!-\!x_{2})\}$ is set to zero when $|g(x_{1}\!-\!x_{2})|\!\leq\!0.05$ to avoid errors stemming from the low signal level.}
\label{Fig6:reflective-object}
\end{figure*}

\subsubsection{Impact of the object longitudinal location}

Finally, we show the effect of moving a $w\!=\!0.5$-mm-wide object along the longitudinal axis $z$ in Fig.~\ref{Fig5:varying-d1}. We hold the total distance between the source and detection plane $d$ fixed and increase $d_{1}$. As the object approaches the detection plane (descreasing $d_{2}$) while remaining on the optical axis ($x_{0}\!=\!0$), the nulls in $|g(x_{1}\!-\!x_{2})|$ move to smaller values $x_{1}\!-\!x_{2}$. In other words, the effect of reducing $d_{2}$ for fixed $w$ is similar to that of increasing $w$ for fixed $d_{2}$. Indeed, from the van Cittert-Zernike theorem, we expect the width of the coherence function to be related to $w/d_{2}$; that is, the angle subtended by the object at the detection plane. Once again, although the van Cittert-Zernike theorem is usually applied to cases where the source is completely incoherent, it is still expected that the general features will apply to a partially coherent field.

\subsection{Coherence function due to a reflecting object}

We now consider reconfiguring the optical arrangement to accommodate the object in reflection mode. Only light reflecting from the object reaches the detection plane [Fig.~\ref{Fig6:reflective-object}]. We collect light that is scattered from the object. The reflective objects were rectangular sections of mirrors of varying widths $w$. Because light is obliquely incident on the object, the effective size is reduced by the cosine of the incidence angle (the angle between incident and reflected light is $\approx\!16^{\circ}$). We expect that if the reflective object size is very large, then light from the source reaches the detection plane with little modification, so that the measured coherence function approaches that of the source [Fig.~\ref{Fig2:source-char}(a)]. \textit{Reducing} the reflective object size, on the other hand, is expected to affect the measured coherence by increasing the width of the coherence function (which is in inverse proportion to the size of the secondary source).

The measurement results are presented in Fig.~\ref{Fig6:reflective-object}. We measured the coherence function while varying the width $w$ of the reflective objects from 0.5~mm to 1~mm. The object is placed on the optical axis of the source and its axial distance from the source and detection plane are held fixed. The measured coherence function does not display nulls or a significant side lobe in contrast to the case of intercepting objects. Indeed, the measured $|g(x_{1}\!-\!x_{2})|$ from the secondary source resembles that of the primary source except from the increased coherence width as the size of the object is reduced. The phase $\mathrm{Arg}\{g(x_{1}\!-\!x_{2})\}$ is flat throughout. We expect that reducing the size of the object further will ultimately introduce nulls in the coherence function and $\pi$-phase jumps in its phase distribution.

\section{Discussion and Conclusion}

\begin{figure*}[t!]
\centering
\includegraphics[scale=.42]{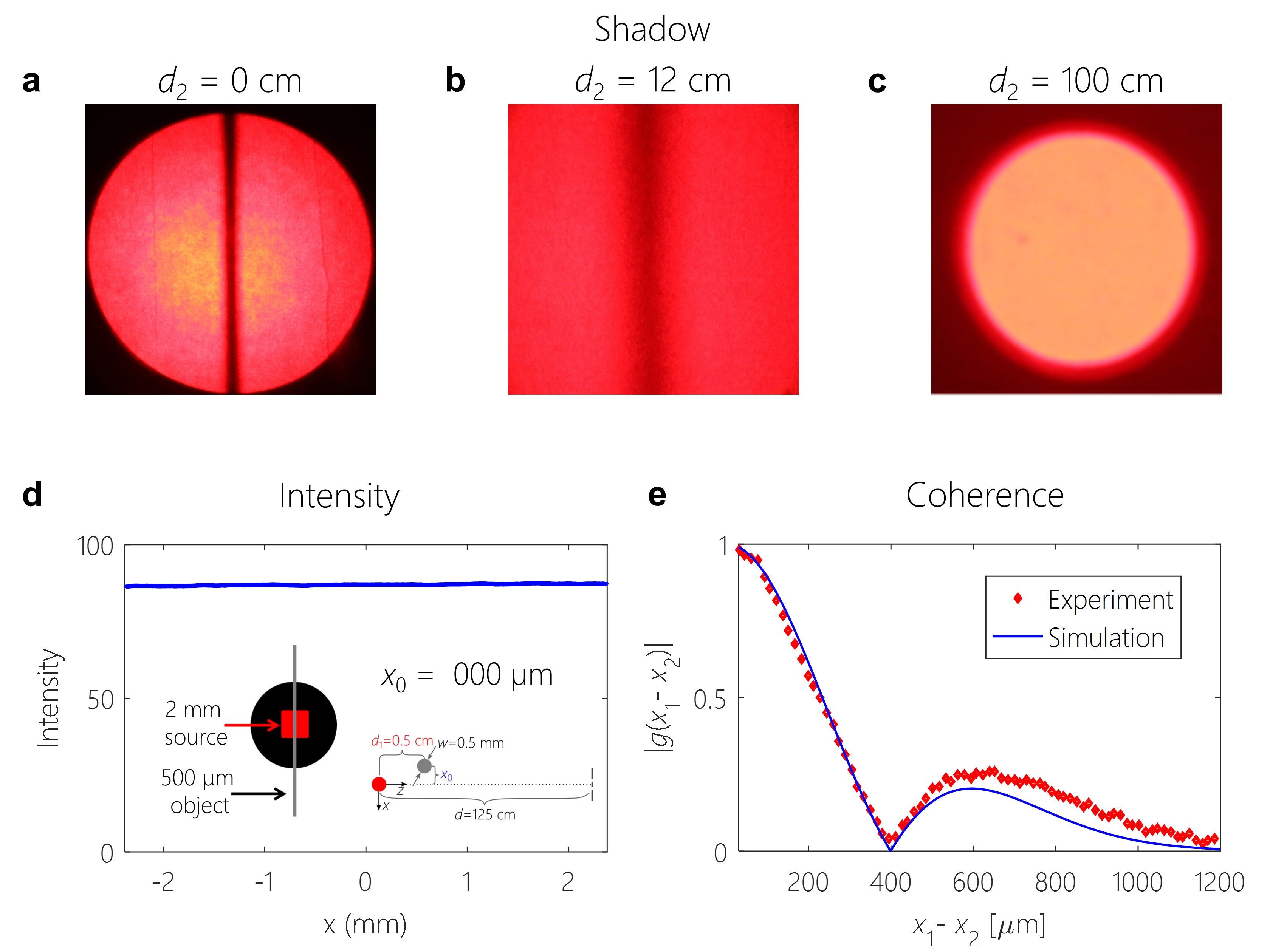}
\caption{Video showing the variations in intensity and coherence as an object is moved transversely in front of the light source (see Visualization 1). Each frame in the video corresponds to the object displaced to a different transverse position $x_{0}$. (a) Image of the shadow cast by a 0.5-mm-wide metal wire (object) moving in front of the LED source in the direct vicinity of the object ($d_{2}\!\approx\!0$). A geometric cast shadow is clear. Image size is $25\!\times\!25$~mm$^2$. (b) Image of the shadow cast by the object at $d_{2}\!\approx\!12$~cm. The shadow is now blurred. Image size is $25\!\times\!25$~mm$^2$. (c) Image of the shadow cast by the object at $d_{2}\!\approx\!100$~cm. The shadow is no longer discernible. Image size is $25\!\times\!25$~mm$^2$. (d) Measured intensity profile $I(x_{1};d_{2}\!=\!100~\mathrm{cm})$ along the center of the beam in (c) for different transverse positions $x_{0}$. Inset shows a schematic of the setup and the transverse motion of the object in the range $-0.5\!\leq\!x_{0}\!\leq\!0.5$~mm. (e) Measured $|g(x_{1}\!-\!x_{2})|$ corresponding to the displaced positions of the object. When the object is on the optical axis, nulls develop in the coherence function. Lifting the ambiguity with respect to the position of the object on the left or the right of the optical axis requires measuring the phase $\mathrm{Arg}\{g(x_{1}\!-\!x_{2})\}$ [see Fig.~\ref{Fig4:varying-x0}].}
\label{Fig7:coh_vs_intensity}
\end{figure*}

The far-field intensity distribution of incoherent or partially coherent light scattering from an object -- in absence of a lens -- diffractively spreads and no longer displays sufficient information to reconstruct the object. Nevertheless, the complex spatial coherence function retains information that reveals the size and location of a thin obstructing or reflective object. Despite the large extent of the intensity distribution in the far field, only a limited area of the field need be examined -- on the order of the width of the spatial coherence at the detector plane. The findings reported here are highlighted by the example shown in Fig.~\ref{Fig7:coh_vs_intensity}, where a metal wire of width $w\!=\!0.5$~mm is scanned in front of the source ($d_{1}\!=\!0.5$~cm). In the immediate vicinity of the object, a geometric shadow is cast ($d_{2}\!=\!0$). At a larger distance ($d_{2}\!=\!12$~cm), the shadow is blurred. At an even larger distance ($d_{2}\!=\!100$~cm), the shadow is no longer discernible. Despite the almost constant far-field intensity profile, measuring $g(x_{1}\!-\!x_{2})$ reveals a clear signature for detecting the change in the transverse position of the object. Note that a moving object, is not required for the procedure to work and the position of a stationary object can also be identified. The availability of high-speed DMDs and sensors can help make this strategy a real-time detection scheme.

This scenario is characterized by significant priors: it is known that there is only one object in the path of the field, which is characterized by only its size and location. In this case, only the anti-diagonal $G(x_{1}\!-\!x_{2})$ of the full coherence function $G(x_{1},x_{2})$ is needed. However, if such a prior is not available -- for example, multiple objects may lie in the path of the field, then it becomes necessary to measure $G(x_{1},x_{2})$. For `sparse' scenes formed of well-separated objects, back-propagation of the coherence function with the Hermitian conjugate of the Fresnel diffraction operator may help reconstruct the full scene. We will present our results on that scheme elsewhere \cite{ElHalawany17}. Finally, we have considered only scalar fields here, but this coherence-measurement scheme may be extended to vector fields by adopting a more general description of vector coherence \cite{Kagalwala13NP,Abouraddy14OL,Kagalwala15SR}.

In this paper, we examined the effect of parameters of a 1D object placed in the path of a partially coherent field -- whether obstructing  or reflecting it -- on the spatial coherence function measured in the far field without a lens. Although the information in the intensity profile is blurred by diffraction, the coherence function retains useful information. In addition, the coherence function remains relatively localized, thus necessitating measurements over only a small area in the far field -- potentially smaller than the physical size of the object itself. Certain features of the coherence function can be exploited to track objects. The size and the location of the object can be retrieved from the coherence function measured rapidly by wavefront sampling via dynamically reconfigurable double slits implemented by a DMD. The rapid refresh rate of such devices may allow for real-time lensless tracking of a moving object in the far field.

\section*{Funding}
Defense Advanced Research Projects Agency (DARPA), Defense Science Office under contract HR0011-16-C-0029.

\section*{Acknowledgments} 
We thank A. Mahalanobis for useful discussions and for bringing Ref.~\cite{George97OC} to our attention. We also grateful to A. Tamasan and A. Dogariu for helpful discussions.

\clearpage
\bibliography{main}
\end{document}